\documentclass[twocolumn,showpacs,aps,floatfix]{revtex4}

\usepackage{amsmath2000}
\usepackage{graphicx}
\usepackage{dcolumn}
\usepackage{bm}

\begin{document}
\preprint{ND Atomic Theory 2002-3}
\title{Two-photon E1M1 decay of $2\, ^{3}\!P_{0}$
states in heavy heliumlike ions}
\author{I. M. Savukov}
 \email{isavukov@nd.edu}
 \homepage{http://www.nd.edu/~isavukov}
\author{W. R. Johnson}
 \email{johnson@nd.edu}
 \homepage{http://www.nd.edu/~johnson}
\affiliation{Department of Physics, 225 Nieuwland Science Hall\\
University of Notre Dame, Notre Dame, IN 46566}

\date{\today}
\begin{abstract}

Two-photon E1M1 transition rates are evaluated for heliumlike ions 
with nuclear charges in the range $Z$ = 50--94. The two-photon rates 
modify previously published lifetimes/transition rates 
of $2\, ^{3}\! P_{0}$ states. 
For isotopes with nuclear spin $I\neq 0$, where hyperfine quenching dominates the
$2\, ^{3}\! P_{0}$ decay, two-photon contributions are significant;
for example, in heliumlike $^{187}$Os the two-photon correction is 
3\% of the total rate.
For isotopes with $I= 0$, where the $2\, ^{3}\! P_{0}$ decay is unquenched, 
the E1M1 corrections are even more important reaching 60\% for Z=94.
Therefore, to aid in the interpretation of experiments on hyperfine
quenching in heliumlike ions and to provide a more complete database for
unquenched transitions, a knowledge of E1M1 rates is important. 
\end{abstract}
\maketitle
\section{Introduction}

In recent years, accurate calculations of decay rates of 
$2\,^{3}\! P_{0}$ states of heliumlike ions have been performed 
for nuclear charges $Z \leq  94$,
with \cite{e1m1:97,e1m1:92} and without \cite{e1m1:95} consideration of 
hyperfine quenching. 
Single photon decay of the $2\,^{3}\! P_{0}$ state to the ground 
state is strictly forbidden by angular momentum selection rules
and only a weak E1 decay to the $2\,^{3}\! S_{1}$ state 
or an even weaker M1 decay to the $2\,^{3}\! P_{1}$  state
is possible for isotopes with nuclear spin $I=0$. For isotopes
with $I\neq 0$,  
weak transitions induced by the hyperfine interaction also are possible. 
Until now, corrections to $2\,^{3}\! P_{0}$ decay rates 
associated with two-photon E1M1 decay have been neglected.
For low-$Z$ ions, those corrections are extremely tiny, scaling 
as $Z^{12}$; however, for ions such as U$^{+90}$, 
the E1M1 decay rate is comparable to 
the single-photon rate as shown by \citet{e1m1:85}. 
Two-photon E1M1 rates are also expected to be significant
for heavy two-electron ions other than U$^{+90}$
and calculations are needed. To date, the only ion with $Z>50$ for which
a measurement of hyperfine quenching has been made is heliumlike gadolinium
\cite{e1m1:92}; for the isotope $^{155}$Gd, the 
 E1M1 correction to the lifetime is only  0.08\%.

We evaluate two-photon E1M1 transition rates 
for $2\, ^{3}\!P_{0}$
states of heliumlike ions with $Z = $ 50--94. The E1M1
rate is a smooth function of $Z$ that follows approximately the $Z^{12}$ law 
predicted by \citet{e1m1:85}. 
Revised lifetimes of $2\, ^{3}\!P_{0}$ states for isotopes with $I\neq 0$,
where hyperfine quenching occurs, are also given. 
Additionally, we give $^{3}P_{0}\rightarrow 1^{1}S_{0}$ rates for 
ions with $Z=50-94$ having $I=0$.
Although E1M1 transitions are very interesting for 
relativistic theories, because of the very large
negative-energy contributions to the transition amplitudes,
direct measurements of E1M1 transitions have not yet been made.

\section{Theory}

\subsection{Two-photon E1M1 transition rate}

The differential transition probability for the
two-photon E1M1 decay, after summation over the
photon polarization states and integration 
over photon angles, takes the form 
\begin{equation}
dw_{FI}=\frac{2}{27\pi }\alpha ^{8}\omega _{1}^{3}\omega _{2}^{3}d\omega
_{1}\left(\left| M(1,2)\right| ^{2}+\left|M(2,1)\right| ^{2}\right)\text{,}
\label{eq1}
\end{equation}
where the photon frequencies are related by energy
conservation, $\omega _{1}+\omega _{2}=E_{f}-E_{i}$ and the two-photon
matrix element $M(1,2)$ is given by, 
\begin{eqnarray}
\lefteqn{M(1,2)=\sum_{n}\frac{\left\langle i\left| E_{1}(1)\right| n\right\rangle
\left\langle n\left| M_{1}(2)\right| f\right\rangle }{E_{n}+\omega _{1}-E_{i} }
}\hspace{3em} \nonumber \\
&& +
\frac{\left\langle i\left| M_{1}(2)\right| n\right\rangle \left\langle
n\left| E_{1}(1)\right| f\right\rangle }{E_{n}+\omega _{2}-E_{i}}\text{.}
\end{eqnarray}
In this equation,  $i$, $f$, and $n$ designate initial, final, and
intermediate states. Permutation of arguments in the matrix element $M(1,2)$
means that $\omega _{1}$ and $\omega _{2}$ should be interchanged in the
denominators and in the arguments of dipole matrix elements. 
Compared to the expression for the 2E1 transition rate given
in Ref.~\cite{e1m1:97a}, there is the extra factor $\alpha ^{2}/4$, 
owing to differences between magnetic and electric
dipole matrix elements.

The matrix element $M(1,2)$ is gauge invariant. The
difference between the length and velocity forms of the electric-dipole
moment in a local potential is $\left\langle i\left| \Delta E_{1}(\omega
)\right| j\right\rangle =a(E_{i}-E_{j}-\omega )\left\langle i\left| \chi
(\omega )\right| j\right\rangle $, where $\chi (\omega )$ is the gauge
operator defined in Ref.~\cite{e1m1:95} and $a$ is a constant of proportionality. 
Using this relation, the completeness of
the basis, and the commutativity of $M_{1}(\omega _{2})$ and $\chi (\omega
_{1})$, we find that the length-velocity difference is for the first term
\begin{equation*}
\frac{\left\langle i\left| \Delta E_{1}(\omega _{1})\right| n\right\rangle
\left\langle n\left| M_{1}(\omega _{2})\right| f\right\rangle }{%
E_{i}-E_{n}-\omega _{1}}=a\left\langle i\left| \chi (\omega _{1})M_{1}(\omega
_{2})\right| f\right\rangle
\end{equation*}
and for the second term 
\begin{equation*}
\frac{\left\langle i\left| M_{1}(\omega _{2})\right| n\right\rangle
\left\langle n\left| \Delta E_{1}(\omega _{1})\right| f\right\rangle }{%
E_{i}-E_{n}-\omega _{2}}=-a\left\langle i\left| M_{1}(\omega _{2})\chi
(\omega _{1})\right| f\right\rangle \text{.}
\end{equation*}
Adding the two terms, we obtain $\Delta M(1,2)=0$.  The numerical calculations
discussed below give precise seven-digit gauge invariance, 
in harmony with the theoretical  prediction.

After summation over magnetic substates, the two-photon matrix element can
be further simplified, 
\begin{multline}
 M(1,2)=\frac{1}{\left[ J_{i}\right] }
\left| \sum_{n}\frac{\left\langle i\left| \left| E_{1}(\omega _{1})\right|
\right| n\right\rangle \left\langle n\left| \left| M_{1}(\omega _{2})\right|
\right| f\right\rangle }{E_{i}-E_{n}-\omega _{1}} \right. \\
+\left. \frac{\left\langle i\left|
\left| M_{1}(\omega _{2})\right| \right| n\right\rangle \left\langle n\left|
\left| E_{1}(\omega _{1})\right| \right| f\right\rangle }{E_{i}-E_{n}-\omega
_{2}}\right| ^{2} .\label{eqm12g}
\end{multline}
Here $\left[ J_{i}\right]=2J_{i}+1$ is the degeneracy of the initial state. 
The quantity $M(1,2)$ is expressed in
terms of reduced dipole matrix elements, which can be easily calculated.
The units of the reduced magnetic-dipole matrix element, which sometimes
differ in literature, are specified here by its relation to the transition
rate 
\begin{equation*}
A_{if}=\frac{2.69735\times 10^{13}}{\lambda ^{3}}\frac{S_{M1}}{g_{i}}\;s^{-1}%
\text{,}
\end{equation*}
where $S_{M1}=\left| \left\langle i\left| \left| M_{1}(\omega )\right|
\right| f\right\rangle \right| ^{2}$ , $g_{i}$ is degeneracy of the initial
state, and $\lambda $ is wavelength in \AA .

We consider two-photon decay of the $^{3}\!P_{0}$ level in the
single-configuration approximation, similar to Ref. ~\cite{e1m1:85}. 
The initial
state is $|i\rangle = |1s2p_{1/2}\, [0]\,\rangle $ , 
the intermediate states are $| n \rangle = |1sns\,[1]\,\rangle $ ($n\neq 1$) or
 $|1snd\, [1]\,\rangle$, and the final
state is the ground state $|f\rangle = | 1s^{2}\,[0]\,\rangle$. 
Substituting these states into (Eq.~\ref{eqm12g}) 
we arrive at the expression for the two-photon amplitude
used in our calculations:
\begin{widetext} 
\begin{eqnarray}
 M(1,2) &=&\!\! \sum_{n\neq 1} 
\frac{\langle 1s_{1/2}2p_{1/2}[0]\, \| E_1(\omega_1)\| 1s_{1/2}ns_{1/2}[1]\, \rangle\, 
\langle 1s_{1/2}ns_{1/2}[1]\, \| M_1(\omega _2)\| 1s_{1/2}^2[0]\, \rangle }
{ E_i-E_{1sns_{1/2}[1]}-\omega _1} \nonumber \\
&+& \sum_{n} 
\frac{\langle 1s_{1/2}2p_{1/2}[0]\, \| E_1(\omega_1)\| 1s_{1/2}nd_{3/2}[1]\, \rangle\, 
\langle 1s_{1/2}nd_{3/2}[1]\, \| M_1(\omega _2)\| 1s_{1/2}^2[0]\, \rangle }
{ E_i-E_{1snd_{3/2}[1]}-\omega _1} \nonumber \\
&+&\sum_{n}
\frac{\langle 1s_{1/2}2p_{1/2}[0]\, \| M_1(\omega _2)\| 1s_{1/2}np_{1/2}[1]\, \rangle\,
 \langle 1s_{1/2}np_{1/2}[1]\| E_1(\omega_1)\| 1s_{1/2}^2[0]\, \rangle }
{E_i-E_{1snp_{1/2}[1]}-\omega _2}\nonumber \\
&+&\sum_{n}
\frac{\langle 1s_{1/2}2p_{1/2}[0]\, \| M_1(\omega _2)\| 1s_{1/2}np_{3/2}[1]\, \rangle\,
 \langle 1s_{1/2}np_{3/2}[1]\| E_1(\omega_1)\| 1s_{1/2}^2[0]\, \rangle }
{E_i-E_{1snp_{3/2}[1]}-\omega _2}. \label{eqq}
\end{eqnarray}
\end{widetext}
With the aid of the expressions for the reduced two-electron matrix elements given 
explicitly in \cite[p.\ 264]{e1m1:95}, we can rewrite (Eq.~\ref{eqq})  
in terms of single-electron reduced matrix elements, leading to
a result in agreement with that given by \citet{e1m1:85}. 

\subsection{Line shape}

One problem of two-photon expressions is distinguishing between ``pure'' and
cascade two-photon processes. The separation is needed for calculations of a
total rate since the cascade process is already included in a single-photon
decay rate but the pure two-photon process is not. (Actually, this provides
a good definition for a cascade process.) Although for the off-resonance
contribution the cascade process is negligible, near the resonance it
dominates. In \cite{e1m1:85}, the subtraction of quadratic resonant terms was
used to calculate the pure two-photon contribution. Since some details are
not clear, we describe in the appendix a subtraction procedure in more detail. We
define a cascade process as the quadratic resonant term with a fixed
frequency in the numerator (this was not stated in \cite{e1m1:85} and is important).
This definition is reasonable because it gives a Lorentzian profile and
after integration over frequencies of the resonant photon the expected
single-photon rate, provided one resonant channel dominates as is the case.
Subtracting this cascade process from the total two-photon contribution, we can define the pure two-photon spectrum. Near resonance portion of the spectrum becomes
closely antisymmetric and upon symmetric integration will contribute
insignificantly. Resonant contributions can then be discarded placing grid
points far away from resonant regions and the pure two-photon contribution
can be trivially evaluated. In the Appendix, we analyze the line
shape of the pure two-photon spectrum near the strong resonance (there is also
a second, weaker resonance but its contribution is less important).

\section{Numerical method}

Since we wish to consider highly-charged ions, we base our
calculations on the Dirac equation. To treat small correlation effects,
of order $1/Z$, we use three starting potentials: Coulomb potential with 
$Z=Z_\text{ion}$, Coulomb potential with $Z=Z_\text{ion}-1$, 
and the model potential 
$V(r) = -Z/r + v_{0}(1s,r)$, where $v_{0}(1s,r)$ is the electrostatic
potential of a single $1s$ electron.  
Among the three,
the model potential takes into account correlations most completely, 
in that experimental energy levels are closely reproduced. 
Therefore, in our final compilation of
two-photon rates, the model potential results are used.

To carry out the sum over intermediate states, for each of the three
potentials we use a B-spline basis set consisting of 40 positive-energy and 40
negative-energy basis orbitals for each
angular momentum state constrained to a cavity of radius 
$R=92/Z_\text{ion}\ a_0$. 
The completeness of this
basis is reflected by precise gauge invariance: for example, in the Coulomb
potential $Z=92$ for heliumlike uranium, the reduced two-photon matrix
elements in the length and the velocity forms are $M(1,2)_{l}=2.9395608\times
10^{-6}$ and $M(1,2)_{v}=2.9395610\times 10^{-6}$. Negative energy
contributions, which almost for 2E1 transitions are small, are
important in the case of E1M1 transitions.

Since four of the terms in Eq.~(\ref{eqq}) have resonant energy denominators
and the two-photon transition rate is extremely sensitive to photon
frequencies ($\sim\omega _{1}^{3}\omega _{2}^{3}$), 
we replace the approximate lowest-order energies by
accurate two-electron energies in these four terms: 
$E_i \rightarrow E(1s^{2\;1}S_{0})$, 
$E_{1s2s[1]} \rightarrow E(1s2s^{\;3}S_{1})$,
$E_{1s2p_{1/2}[1]} \rightarrow E(1s2p^{\;3}P_{0})$, 
and $ E_{1s2p_{3/2}[1]} \rightarrow E(1s2p^{\;1}P_{1})$,
where the accurate energies $E(1snl^{\;1,3}L_{J})$ are taken from \citet{e1m1:95a}.
Similar replacements were used by~\citet{e1m1:85} in his
calculations of the two-photon rate for U$^{90+}$.

\section{Results}

Using the procedure described in previous section, we compute the two-photon
transition rates according to the Eq(\ref{eq1}) and after subtraction resonances
integrate the differential rate to obtain the total two-photon rate. Since
the behavior of the rates with $Z$ is very smooth, we performed calculations
only for eleven different ions. Table~\ref{Table1} summarizes our results for total
transition rates. The three potentials, discussed above, give results in
close agreement, but the model potential for which length-velocity agreement
ranges 0.4-1.5\% and so does approximately accuracy should be considered the
most accurate. (The precise gauge-invariance is achieved when denominators are not modified in all three cases.) For U$^{90+}$our Coulomb $Z$ potential result ($5.567\times
10^{9}$ s$^{-1}$) agrees well with Drake's ~\cite{e1m1:85} ($5.60\times 10^{9}$ s$^{-1}$)
obtained in the same potential. There is also small E2M2 contribution equal
to ($0.025\times 10^{9}$ s$^{-1}$) which was calculated by~\citet{e1m1:85} for U$%
^{90+} $. We will neglect it in all our calculations.

\begin{figure}
\centerline{\includegraphics[scale=0.6]{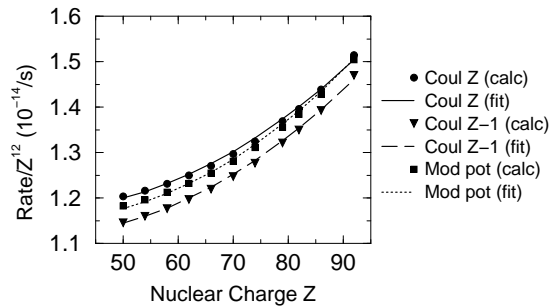}} 
\caption{\label{Fig2} Polynomial interpolations of 
$Z$-dependences for total transition rates calculated in the three starting potentials.}
\end{figure}

\begin{table}
\caption{E1M1 total transition rates for heliumlike ions in the range $%
Z=50-92$.  The rates were calculated using three
different potentials: the Coulomb $Z$, the Coulomb $Z-1$, and a model
potential $V=-Z/r+v_{0}(1s,r)$. In the model potential case, the
difference between results obtained in length and velocity forms are
also shown as a measure of accuracy of calculations.}
\label{Table1}
\begin{tabular}{c c c c c c}
\hline
\hline
\multicolumn{1}{c}{$Z$}& 
\multicolumn{1}{c}{}&
\multicolumn{1}{c}{Coul $Z$} &
\multicolumn{1}{c}{Coul $Z$-1} &
\multicolumn{1}{c}{MP} &
\multicolumn{1}{c}{(L-V)/L} \\ 
\hline
50 & &2.939[6] & 2.798[6] & 2.888[6] & 1.7\% \\
54 & &7.476[6] & 7.130[6] & 7.354[6] & 1.3\% \\
58 & &1.784[7] & 1.706[7] & 1.758[7] & 1.0\% \\
62 & &4.032[7] & 3.862[7] & 3.976[7] & 0.8\% \\
66 & &8.686[7] & 8.335[7] & 8.575[7] & 0.7\% \\
70 & &1.795[8] & 1.725[8] & 1.774[8] & 0.5\% \\
74 & &3.574[8] & 3.442[8] & 3.536[8] & 0.5\% \\
79 & &8.090[8] & 7.808[8] & 8.014[8] & 0.5\% \\
82 & &1.291[9] & 1.247[9] & 1.279[9] & 0.4\% \\
86 & &2.355[9] & 2.280[9] & 2.337[9] & 0.4\% \\
92 & &5.567[9] & 5.403[9] & 5.530[9] & 0.5\% \\
\hline
\end{tabular}
\end{table}
In the table, we provide one extra digit beyond accuracy for better
interpolation. For convenience, we also give the interpolation polynomials
that fit our calculated rates with sufficient accuracy. In the case of the
Coulomb $Z$ potential the polynomial is $Z^{12}(1.28-6.4\times
10^{-3}Z+9.60\times 10^{-5}Z^{2})$; in the case of the Coulomb $Z-1$
potential the polynomial is $Z^{12}(1.22-6.4\times 10^{-3}Z+9.8\times
10^{-5}Z^{2})$; and in the case of the model potential the polynomial is $%
Z^{12}(1.29-7.73\times 10^{-3}Z+1.10\times 10^{-4}Z^{2})$. The rates in all
three cases as well as the fitting polynomials are shown in Fig.~\ref{Fig2}.

\begin{figure}[b]
\centerline{\includegraphics[scale=0.6]{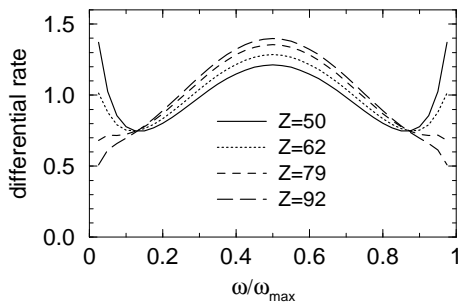}}
\caption{\label{Fig3} Differential rate for two-photon decay}
\end{figure}

The upper curve corresponds to the Coulomb $Z$ potential, the lower curve to
the Coulomb $Z-1$ potential, and the middle one to the model potential. In
addition to total rates we also show in (Fig.~\ref{Fig3}) the frequency distribution
of the differential two-photon transition rate for $Z=92$, $Z=79$, $Z=62$,
and $Z=50$. In the case of uranium, $Z=92$, the shape resembles very closely
the distribution given by Drake~\cite{e1m1:85}. The continuous evolution with $Z$
can be seen: the maximum at the center monotonously decreases with $Z$, but
the wings are rapidly developing for lower $Z$ ions.

In the case of $Z=50$, the wings are large; therefore we studied
the line shape more carefully with a refined scale. The discussion of the
behavior near the origin, where resonances are located, is given in the
Appendix. From this discussion follows that the integrated
contribution from that region is not much different from that of any other
part of the spectrum. Since we discovered that E1M1 corrections are
significant for ions in the range Z=50-94, we also correct previously
calculated values of lifetimes and transition rates for those ions. 
Tables~\ref{Table2} and~\ref{Table3}
show previously calculated and corrected rates. 
Table~\ref{Table2} gives revised
lifetimes for isotopes that have hyperfine induced transitions. 
We show only
the cases where modification changes quoted digits. Table~\ref{Table3} shows transition rates
without hyperfine mixing, when corrections owing to E1-M1 transitions are particularly large.
\begin{table}
\caption{E1M1 two-photon corrections to lifetimes $\tau$ (ps) and E1M1 rates $A_{E1M1}$ (1/ns) 
for ions with hyperfine induced decays.}
\label{Table2} 
\begin{tabular}{cccccc}
\hline
\hline
\multicolumn{1}{c}{Ion} &
\multicolumn{1}{c}{$Z$} &
\multicolumn{1}{c}{$\tau_\text{old}$}&
\multicolumn{1}{c}{$\tau_\text{new}$}&
\multicolumn{1}{c}{$A_{E1M1}$}& 
\multicolumn{1}{c}{\% change} \\ 
\hline
$^{155}$Gd & 64 & 13.57 & 13.56 & 0.059 & 0.08\% \\ 
$^{169}$Tm & 69 & 8.487 & 8.476 & 0.149 & 0.13\% \\ 
$^{171}$Yb & 70 & 1.917 & 1.916 & 0.178 & 0.03\% \\ 
$^{173}$Yb & 70 & 2.113 & 2.112 & 0.178 & 0.04\% \\ 
$^{177}$Hf & 72 & 1.603 & 1.602 & 0.253 & 0.04\% \\ 
$^{179}$Hf & 72 & 2.556 & 2.554 & 0.253 & 0.06\% \\ 
$^{183}$W  & 74 & 25.54 & 25.31 & 0.355 & 0.91\% \\ 
$^{187}$Os & 76 & 59.83 & 58.11 & 0.496 & 2.97\% \\ 
$^{189}$Os & 76 & 1.546 & 1.545 & 0.496 & 0.08\% \\ 
$^{191}$Ir & 77 & 25.07 & 24.71 & 0.584 & 1.46\% \\ 
$^{193}$Ir & 77 & 21.70 & 21.43 & 0.584 & 1.27\% \\ 
$^{195}$Pt & 78 & 0.9433& 0.9430& 0.686 & 0.06\% \\ 
$^{197}$Au & 79 & 23.04 & 22.62 & 0.805 & 1.86\% \\ 
$^{199}$Hg & 80 & 1.248 & 1.247 & 0.943 & 0.12\% \\ 
$^{201}$Hg & 80 & 1.782 & 1.779 & 0.943 & 0.17\% \\ 
$^{203}$Tl & 81 & 0.1215& 0.1210& 1.10  & 0.01\% \\ 
$^{205}$Tl & 81 & 0.1192& 0.1190& 1.10  & 0.01\% \\ 
$^{207}$Pb & 82 & 0.8357& 0.8350& 1.29  & 0.11\% \\ 
$^{209}$Bi & 83 & 0.0041& 0.0040& 1.50  & 0.00\% \\ 
$^{223}$Ra & 88 & 5.079 & 4.999 & 3.14  & 1.60\% \\ 
$^{229}$Th & 90 & 1.955 & 1.939 & 4.18  & 0.82\% \\ 
$^{235}$U  & 92 & 2.623 & 2.585 & 5.54  & 1.45\% \\ 
$^{239}$Pu & 94 & 3.707 & 3.610 & 7.29  & 2.70\% \\ 
\hline
\end{tabular}
\end{table}

\begin{table}
\caption{E1M1 two-photon corrections to transition rates $A$ (1/s) for ions 
without hyperfine induced decays.}
\label{Table3} 

\begin{tabular}{ccccc}
\hline
\hline
\multicolumn{1}{c}{$Z$} &
\multicolumn{1}{c}{$A_\text{old}$} &
\multicolumn{1}{c}{$A_\text{E1M1}$} &
\multicolumn{1}{c}{$A_\text{new}$} &
\multicolumn{1}{c}{\% change} \\ 
\hline
50 & 1.363[9] & 2.87[6] & 1.366[9] & 0.21\% \\
51 & 1.439[9] & 3.66[6] & 1.443[9] & 0.25\% \\
52 & 1.519[9] & 4.63[6] & 1.524[9] & 0.30\% \\
53 & 1.605[9] & 5.84[6] & 1.611[9] & 0.36\% \\
54 & 1.695[9] & 7.33[6] & 1.702[9] & 0.43\% \\
55 & 1.790[9] & 9.16[6] & 1.799[9] & 0.51\% \\
56 & 1.891[9] & 1.14[7] & 1.902[9] & 0.60\% \\
57 & 1.997[9] & 1.42[7] & 2.011[9] & 0.71\% \\
59 & 2.228[9] & 2.16[7] & 2.250[9] & 0.97\% \\
60 & 2.353[9] & 2.66[7] & 2.380[9] & 1.13\% \\
62 & 2.623[9] & 3.97[7] & 2.663[9] & 1.51\% \\
63 & 2.768[9] & 4.84[7] & 2.816[9] & 1.75\% \\
64 & 2.925[9] & 5.87[7] & 2.984[9] & 2.01\% \\
65 & 3.090[9] & 7.11[7] & 3.161[9] & 2.30\% \\
66 & 3.263[9] & 8.59[7] & 3.349[9] & 2.63\% \\
67 & 3.446[9] & 1.03[8] & 3.549[9] & 3.00\% \\
68 & 3.638[9] & 1.24[8] & 3.762[9] & 3.41\% \\
69 & 3.839[9] & 1.49[8] & 3.988[9] & 3.88\% \\
70 & 4.049[9] & 1.78[8] & 4.227[9] & 4.39\% \\
71 & 4.273[9] & 2.12[8] & 4.485[9] & 4.97\% \\
72 & 4.507[9] & 2.53[8] & 4.760[9] & 5.60\% \\
73 & 4.754[9] & 3.00[8] & 5.054[9] & 6.31\% \\
74 & 5.011[9] & 3.55[8] & 5.366[9] & 7.09\% \\
75 & 5.287[9] & 4.20[8] & 5.707[9] & 7.95\% \\
76 & 5.574[9] & 4.96[8] & 6.070[9] & 8.89\% \\
77 & 5.875[9] & 5.84[8] & 6.459[9] & 9.94\% \\
78 & 6.187[9] & 6.86[8] & 6.873[9] & 11.09\% \\
79 & 6.315[9] & 8.05[8] & 7.120[9] & 12.75\% \\
80 & 6.852[9] & 9.43[8] & 7.795[9] & 13.76\% \\
81 & 7.215[9] & 1.10[9] & 8.318[9] & 15.28\% \\
82 & 7.663[9] & 1.29[9] & 8.950[9] & 16.79\% \\
83 & 7.971[9] & 1.50[9] & 9.470[9] & 18.81\% \\
88 & 1.004[10]& 3.14[9] & 1.318[10]& 31.31\% \\
90 & 1.093[10]& 4.18[9] & 1.511[10]& 38.28\% \\
92 & 1.181[10]& 5.54[9] & 1.735[10]& 46.87\% \\
94 & 1.267[10]& 7.29[9] & 1.996[10]& 57.51\% \\
\hline
\end{tabular}
\end{table}
The difference reaches 57.5\% for $Z=94$. (Drake's prediction in uranium~\cite{e1m1:85} was
46\%.) Depending on experimental situation, any ion shown in the table can
be potentially used for the measuring E1M1 transition rate.

\section{Conclusions}

In this paper, we have calculated E1M1 transition rates for heliumlike ions (%
$Z=50-94$). The previous lifetime values are improved with adding E1M1
contributions. The large E1M1 corrections in heavy ions can be tested in future experiments and the E1M1 transition rates can be extracted from experiments to study those essentially relativistic transitions.
\begin{acknowledgments}
The authors thank R. Marrus for suggesting this  problem and acknowledge 
helpful discussions with Professor L. M. Labzowsky. 
This work was supported in part by 
National Science Foundation Grant No.\ PHY-01-39928.
\end{acknowledgments}

\appendix*
\section{Analysis of the line shape}
The two-photon transition matrix element $M(1,2)$ has two
poles: one occurs for $n=2$ in the first term of Eq.~(\ref{eqq}),
when $\omega _1$ equals the 
$2\, ^3\! P_0 \rightarrow 2\, ^3\! S_1$ ($E_1$) transition energy, and the second
occurs for $n=2$ in the third term term, when $\omega _1$ equals the 
$2\, ^3\! P_0 \rightarrow 2\, ^3\! P_1$ ($M_1$) transition energy.
To obtain the correct line
shape, the denominators in each of these pole terms should be modified to include 
widths of the levels: 
$E_{n}\rightarrow E_{n}-i\Gamma /2$. (If the small width of the initial state is
neglected, then $\Gamma =\Gamma _{n}$, the width of
the $n$th level only). The insertion of the imaginary term leads to an almost Lorentzian
profile near the resonance. By subtracting the Lorentzians corresponding to
the cascade processes or simply single-photon decays, we obtain the purely
two-photon decay rate and its spectral distribution.  After subtraction, 
the remaining contribution  to the two-photon rate  near the resonance region becomes
very small when the resonant intermediate level is very narrow, as is the
case for the intermediate  $2\, ^3\!S_1$ level and the resonances can be ignored 
in the total two-photon rate.  To check that resonant contributions are small, we
made fine grid in the resonant region and calculated the resonance
contributions. The cascade contribution is
large and is calculated separately, using more precise methods. A precise rate
for the $2\, ^3\! P_0 \rightarrow 2\, ^3\! S_1$ ($E_1$) transition is given in
Ref.~\cite{e1m1:95}. For comparison with experiment, our two-photon line shape
should be added to the Lorentzian obtained using the accurate line
width and the single-photon transition rate.  In this paper, we will show
only pure two-photon profile, {\it i.e.,}\ what remains after resonances are
subtracted.

\begin{figure}
\centerline{\includegraphics[scale=0.6]{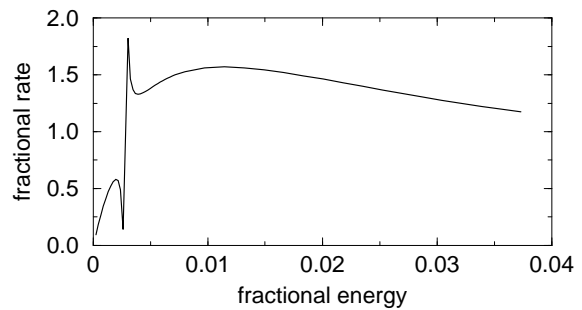}} 
\caption{\label{Fig1} Near-resonant portion of pure two-photon spectrum }
\end{figure}

The analysis of the relative contribution to the two-photon rate from the resonant
region, using limit that the width is very small, gives the following
expression:
\begin{equation*}
I(\omega )\simeq \frac{I_{1}(\omega _{1})}{3\omega _{1}}\left[ \frac{3}{2}%
\ln \left| \omega -1\right| +\omega +\frac{1}{4}\omega ^{2}\right] \text{.}
\end{equation*}
In this formula the unit of frequency is chosen the resonant frequency 
$\omega _{res}$, $I(\omega )$ is the contribution obtained by integration
from zero to $\omega $, $I_{1}(\omega _{1})$ is the contribution with the
symmetric integration from $-\omega _{1}$ to $+\omega _{1}$ with $\omega
_{1}<1$. The integrals $I_{1}(\omega _{1})$ and $I(\omega )$ are defined
below, and the expansion over small parameter $\Gamma $ have been carried
out: 
\begin{equation*}
I_{1}(\omega _{1})=A\int\limits_{1-\omega _{1}}^{1+\omega _{1}}\frac{x^{3}-1%
}{(x-1)^{2}+\left( \frac{\Gamma }{2}\right) ^{2}}dx\simeq A3\omega _{1}\Gamma
\end{equation*}
\begin{eqnarray}
\lefteqn{
I(\omega )=A\int\limits_{0}^{\omega }\frac{x^{3}-1}{(x-1)^{2}+\left( \frac{%
\Gamma }{2}\right) ^{2}}dx
}\hspace{3em} \nonumber \\
&& 
\simeq A\frac{\Gamma }{2}\left[ \frac{3}{2}\ln
\left| \omega -1\right| +\omega +\frac{1}{4}\omega ^{2}\right].
\end{eqnarray}
The illustration of the shape of the near resonance differential rate is
given in Fig.~\protect\ref{Fig1}. Here the case $Z=50$ is considered. The differential
rate near the resonance is very large, of order 10$^{4}$, and is not shown
here. However, owing to a precise cancellation, the total contribution after
integration is quite small.

The symmetric integral in the near resonance region owing to very precise
cancellation provides better numerical accuracy than asymmetric, but far
away from the resonance when $\omega _{1}>1$, instead of the symmetric
integral the asymmetric integral has to be used. We assume that only one
strong resonance near zero where two-photon differential rate is
proportional to $\omega ^{3}$ is present. The integral $I_{1}(\omega _{1})$
is proportional to the frequency $\omega _{1}$ and to the width of the
resonance, and the smaller the width the smaller resonant contribution 
$I(\omega )$ will be. Using the above equations, it can be shown that 
the residual resonant contribution to the two-photon rate 
is insignificant compared to
the smooth continuum contribution.

The shape in the near resonance region can be described approximately by the
following equation
\begin{equation*}
f(x)\approx \frac{1+x+x^{2}}{(x-1)}+background
\end{equation*}
if $(x-1)\gg \Gamma $. The background is the smooth continuum contribution
owing to all terms except the resonant term. There is maximum at $x=0.012$
owing to two-photon continuum for small $x$ behaving like $x^{k}(1-x)^{3}$.
Depending on the position of most contributing levels, $k=1\div 3$ . We show
the case $Z=50$ for which the wings are the most pronounced. At the very
origin, the differential rate rapidly grows $x^{3}$ if $x$ is much less than
the any denominator of the terms with large contributions.

The sharp left edge of the wings is owing to the sharp $\omega ^{3}$
dependence of the differential rate near the origin. The located very close
to the origin subtracted resonances, which have asymmetric shapes and narrow
widths, are shown in Fig.~\ref{Fig1} since their differential rate near the resonance
frequency is very high and needs different scale. But resonant contributions
also decrease so fast that they have no effect on the shown region as well
on the total rate.


\end{document}